\documentclass[preprint]{revtex4-1}

\usepackage{graphicx}
\usepackage{bm}
\usepackage[
colorlinks=true,
linkcolor=blue,
citecolor=blue,
filecolor=blue,
urlcolor=blue]{hyperref}

\usepackage{amsmath,amsfonts,amsgen,amsbsy,amsopn,amstext,amssymb}
\usepackage{epstopdf}
\usepackage{color}

\renewcommand{\i}{i}
\renewcommand{\(}{\left(}
\renewcommand{\)}{\right)}
\renewcommand{\Pr}{\mathrm{Pr}}

\setrefcountdefault{0}

\begin{document}

\preprint{}

\title{Ultra-thin metamaterial for perfect and omnidirectional sound absorption}

\author{N. Jim\'enez}
\email{noe.jimenez@univ-lemans.fr}

\author{W. Huang}
\author{V. Romero-Garc\'ia}
\author{V. Pagneux}
\author{J.-P. Groby}
\affiliation{Laboratoire d'Acoustique de l'Universit\'e du Maine (LAUM) - CNRS UMR 6613, Av. Olivier Messiaen, 72085, Le Mans, France}

\date{\today}

\begin{abstract}
Using the concepts of slow sound and of critical coupling, an ultra-thin acoustic metamaterial panel for perfect and omnidirectional absorption is theoretically and experimentally conceived in this work. The system is made of a rigid panel with a periodic distribution of thin closed slits, the upper wall of which is loaded by Helmholtz Resonators (HRs). The presence of resonators produces a slow sound propagation shifting the resonance frequency of the slit to the deep sub-wavelength regime ($\lambda/88$). By controlling the geometry of the slit and the HRs, the intrinsic visco-thermal losses can be tuned in order to exactly compensate the energy leakage of the system and fulfill the critical coupling condition to create the perfect absorption of sound in a large range of incidence angles due to the deep subwavelength behavior. 
\end{abstract}

\maketitle

The ability to perfectly absorb an incoming wave field in a sub-wavelength material is advantageous for several applications in wave physics as energy conversion \cite{law2005}, time reversal technology \cite{derode1995}, coherent perfect absorbers \cite{chong2010} or soundproofing \cite{mei2012} among others. The solution of this challenge requires to solve a complex problem: reducing the geometric dimensions of the structure while increasing the density of states at low frequencies and finding the good conditions to match the impedance to the background medium.

A successful approach for increasing the density of states at low frequencies with reduced dimensions is the use of metamaterials. Recently, several possibilities based on these systems have been proposed to design sound absorbing structures which can present simultaneously sub-wavelength dimensions and strong acoustic absorption. One strategy to design these sub-wavelength systems consists of using space-coiling structures \cite{cai2014,li2016}. Another way is to use sub-wavelength resonators as membranes\cite{yang2008,mei2012} or Helmholtz resonators (HRs) \cite{merkel2015, malakas2016}. Recently, a new type of sub-wavelength metamaterials based on the concept of slow sound propagation have been used to the same purpose. This last type of metamaterials \cite{leclaire2015,groby2015,groby2016} makes use of its strong dispersion for generating slow-sound conditions inside the material and, therefore, drastically decreasing frequency of the absorption peaks. Hence, the structure thickness becomes deeply sub-wavelength. All of these structures, however, while they bring potentially solutions to reduce the geometric dimensions, face the challenge of impedance mismatch to the background medium.

The interaction of an incoming wave with a lossy resonant structure, in particular the impedance matching with the background field, is one of the most studied process in the field of wave physics \cite{law2005, derode1995, chong2010}. These open systems, at the resonant frequency, are characterized by both the leakage rate of energy (i.e., the coupling of the resonant elements with the propagating medium), and the intrinsic losses of the resonator. The balance between the leakage and the losses activates the condition of critical coupling, trapping the energy around the resonant elements and generating a maximum of energy absorption \cite{bliokh2008}. In the case of transmission systems, degenerate critically coupled resonators with symmetric and antisymmetric resonances should be used to perfectly absorb the incoming energy by trapping the energy in the resonant element, i.e. without reflection neither transmission \cite{piper2014,yang2015}. In the case of purely reflecting system, either symmetric or antisymmetric resonances that are critically coupled can be used to obtain perfect absorption of energy by a perfect trapping of energy around the resonators \cite{ma2014,romero2016}.

\begin{figure*}[htbp]
	\centering
	\includegraphics[width=13cm]{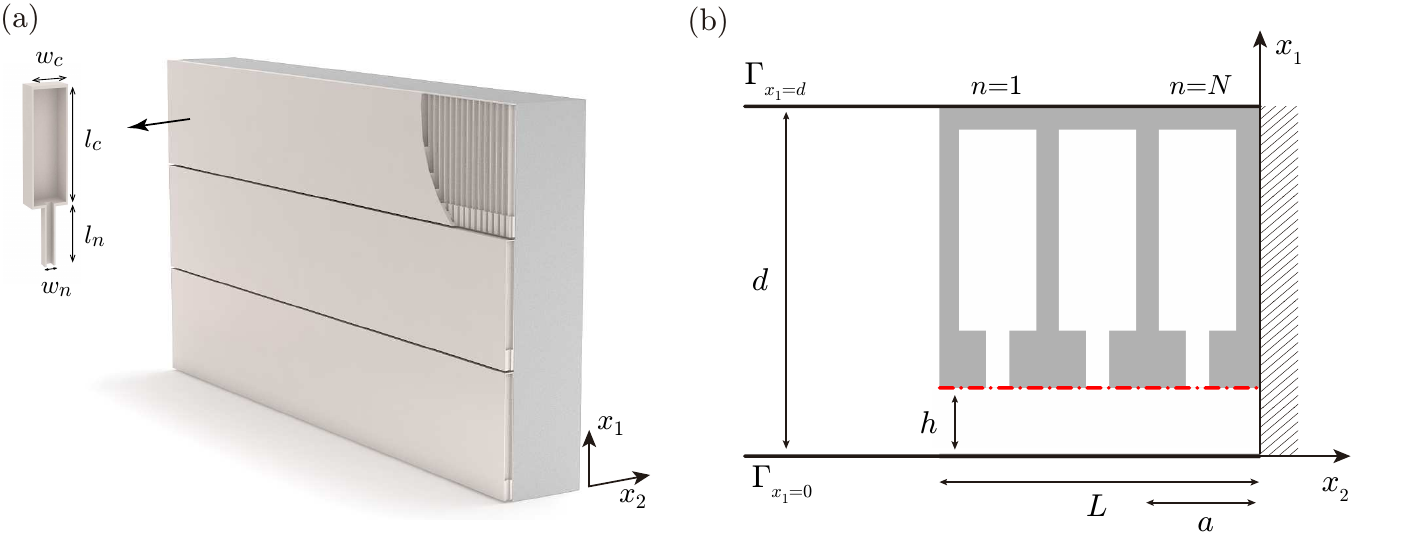}
	\caption{(a) Conceptual view of the thin panel placed on a rigid wall with one layer of square cross-section Helmholtz resonators, $N=1$. (b) Scheme of the unit cell of the panel composed of a set of $N$ Helmholtz resonators. Symmetry boundary conditions are applied at boundaries $\Gamma_{x_1=d}$ and $\Gamma_{x_1=0}$.}
	\label{fig:scheme}
\end{figure*}

In this work, using the concepts of slow sound and critical coupling, we theoretically and experimentally report a prefect and omnidirectional absorbing metamaterial panel with deep sub-wavelength thickness. As shown in Fig.~\ref{fig:scheme}, the system consists of a thin panel perforated with a periodic arrangement of slits, of thickness $h$, with periodicity $d$ along the $x_1$ direction. The upper wall of the slit is loaded by $N$ identical HRs arranged in a square array of side $a$. The HRs, of square cross-section, are characterized by a neck and a cavity widths $w_n$ and $w_c$, and lengths $l_n$ and $l_c$ respectively. The presence of the HRs introduces a strong dispersion in the slit producing slow propagation, in such a way that the resonance of the slit is down shifted: the slit becomes a deep sub-wavelength resonator. The visco-thermal losses in the system are considered in both the resonators and in the slit by using effective complex and frequency dependent parameters \cite{stinson1991}. Therefore, by modifying the geometry, the intrinsic losses of the system can be efficiently tuned and the critical coupling condition can be fulfilled to solve the impedance matching to the exterior medium.

\begin{figure*}[htpb]
	\centering
	\includegraphics[width=11cm]{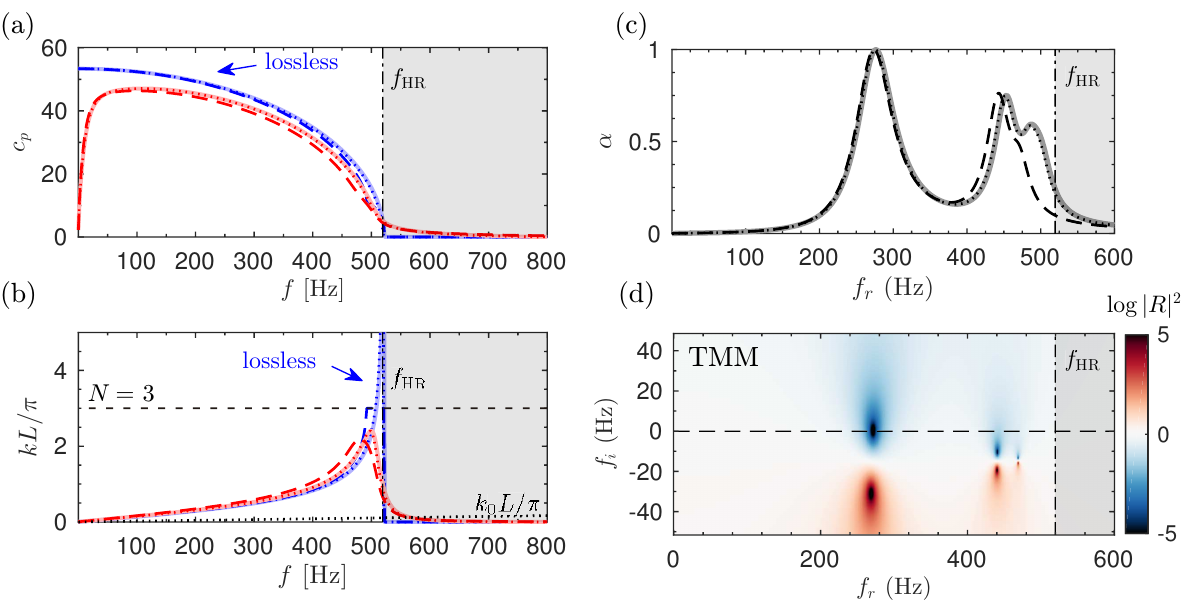}
	\caption{(a) Phase speed for a panel of $N=3$ resonators calculated by full modal expansion (continuous gray), effective parameters (dotted) and TMM (dashed) for the lossless case (blue) and including thermo-viscous losses (red). (b) Corresponding wavenumber, where $k_0$ is the wavenumber in air. (c) Absorption of the panel. The dashed-dotted line marks the resonant frequency of the HRs and the shaded area corresponds to the band-gap. (d) Complex-frequency planes of the reflection coefficient calculated by TMM where $f_r$ and $f_i$ is the real and imaginary part of the complex frequency.}
	\label{fig:N3}
\end{figure*}

We start by analyzing the dispersion properties inside the slit in order to inspect the slow sound behavior. The unit cell of the system considered in this work is shown in Fig. \ref{fig:scheme}(b). Periodic boundary conditions are assumed at boundaries $\Gamma_{x_1=0}$ and $\Gamma_{x_1=d}$. At this stage we have to notice that through this work several theoretical models are used to analyze the structure: a full modal expansion model (MEM), its low frequency approximation that gives the effective parameters, an approach based on the transfer matrix method (TMM) and the finite element method (FEM) (see Supplementary Material for more details of the models). Figure \ref{fig:N3}~(a) shows the real part of the phase velocity in the slit, calculated both in the lossless and lossy cases, for a metamaterial with parameters $h=1.2$ mm, $a=1.2$ cm, $w_n=a/6$, $w_c=a/2$, $d=7$ cm, $l_n=d/3$, and $l_c=d-h-l_n$. Figure \ref{fig:N3}~(b) shows the corresponding dispersion relation, where a band gap can be observed above the resonant frequency of the HRs, $f_\mathrm{HR}$. Due to the presence of this band gap, slow propagation conditions are achieved in the dispersive band below $f_\mathrm{HR}$. In the lossless case, zero phase velocity can be observed for frequencies just below $f_\mathrm{HR}$. Note also that the maximum wavenumber inside the slit is limited by the discreteness to the value $k_\mathrm{max}=\pi N/L$, as shown by the TMM calculations (dashed blue curve in Fig. \ref*{fig:N3}(b)). In the lossy case, the losses limit the minimum value of group velocity \cite{theocharis2014}, but in our system slow sound velocity can be achieved in the dispersive band below $f_\mathrm{HR}$. The average sound speed in the low frequency range is much lower (50 m/s) than the speed of sound in air. Therefore, the collective resonances produced by the array of HRs will be in the deep sub-wavelength regime compared with the length, $L$, of the slit.

The resonant scattering problem analyzed in this work can be characterized by the reflection coefficient. A simple way to obtain this reflection coefficient consists of using the effective bulk modulus, $\kappa_{e}$, and effective density, $\rho_{e}$ of the system \cite{}. The reflection coefficient reads as
\begin{equation}
R_{e}(\theta)=\frac{i Z_{e}'\cot(k_{e}L)-i\omega \Delta l_\mathrm{slit}/\phi_t c_0-1/\cos(\theta)}{i Z_{e}'\cot(k_{e}L)-i\omega \Delta l_\mathrm{slit}/\phi_t c_0+1/\cos(\theta)}\,,
\end{equation}
\noindent where the normalized effective impedance is $Z_{e}'=\sqrt{\rho_{e}\kappa_{e}/\rho_{0}\kappa_{0}}$ ($\rho_0$ and $\kappa_0$ are the density and bulk modulus of air), the effective wavenumber $k_{e}=\omega\sqrt{\rho_{e}/\kappa_{e}}$, $\Delta l_\mathrm{slit}$ is the end correction of the slit accounting for the radiation from the slit to the free space, $\phi_t=h/d$ is the total porosity of the metamaterial, $\omega$ is the angular frequency, $c_0$ the sound speed of air and $\theta$ the angle of incidence. The effective parameters can be obtained in the low frequency approximation of the MEM as
\begin{eqnarray}
\kappa_{e}&=&\frac{{\kappa_s}}{\phi_{t}}{\left[1+\frac{\kappa_s\phi\left(v_c\kappa_n+v_n\kappa_c\right)}{\kappa_n h\left(S_n\kappa_c-v_c\rho_n l_n \omega^2\right)}\right]}^{-1}, \label{eq:ke}\\ 
\rho_{e}&=&\frac{\rho_s}{\phi_{t}}, \, \label{eq:rhoe}
\end{eqnarray}
\noindent where $\rho_s$ and $\rho_n$ are the effective densities of the slit and neck, $\kappa_s$, $\kappa_n$ and $\kappa_c$ are the effective bulk modulus of the slit, neck and cavity respectively, $v_n$ and $v_c$ are the volumes of the neck and cavity of the HRs respectively.

In the complex frequency plane the reflection coefficient has pairs of zeros and poles that are complex conjugates one from another in the lossless case. In the $\exp(-\imath\omega t)$ sign convention, the zeros are located in the positive imaginary plane. The imaginary part of the complex frequency of the poles of the reflection coefficient represents the energy leakage of the system into the free space\cite{romero2016}. Once the intrinsic losses are introduced in the system, the zeros of the reflection coefficient move downwards to the real frequency axis\cite{Romero2016b}. For a given frequency, if the intrinsic losses perfectly balance the energy leakage of the system, a zero of the reflection coefficient is exactly located on the real frequency axis and therefore perfect absorption, $\alpha=1-|R_e|^2=1$, can be obtained. This condition is known as critical coupling \cite{bliokh2008,piper2014,yang2015,ma2014,romero2016,Romero2016b}.

Figure \ref{fig:N3}~(c) shows the absorption predicted by the different models when the geometry of the system has been tuned to introduce the exact amount of intrinsic losses that exactly compensates the energy leakage of the system at 275 Hz for $N=3$. In this situation, as shown in Fig.~\ref{fig:N3}~(d), the lower frequency zero is located on the real axis, leading to a peak of perfect absorption. In addition, as we have $N=3$ resonators, other two secondary peaks of absorption are observed at higher frequencies, e.g. 442 Hz and 471 Hz. Their corresponding zeros are located close to the real axis and, although the critical coupling condition is not exactly fulfilled, high absorption values can be observed at these frequencies. The differences between the several model predictions observed in the absorption coefficient are due to the fact that the effect of the discreteness is not captured by neither the MEM or its effective parameters. MEM leads to an infinite number of zeros and poles close to the resonance, producing the artificial increase of the absorption observed in Fig.~\ref{fig:N3}~(c) near $f_\mathrm{HR}$. Consequently, the prediction of absorption by MEM is accurate only if these zeros do not interact with the real axis or if the number of resonators is considerably increased. The TMM captures this discreteness effect and shows good results in this range of frequencies (400-500 Hz). 

\begin{figure*}[t]
	\centering
	\includegraphics[width=11.5cm]{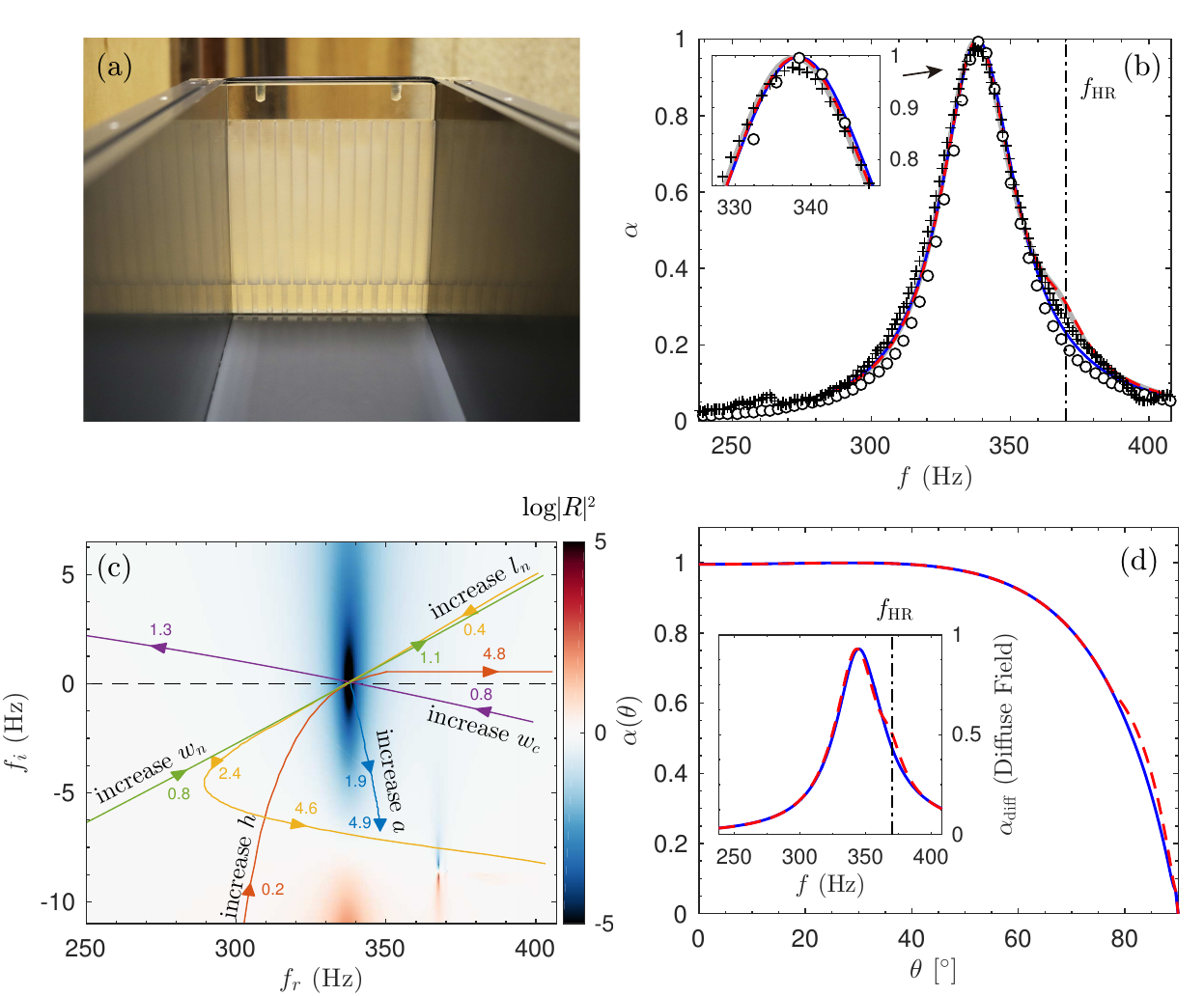}
	\caption{(a) Photograph of the experimental setup with a vertical unit cell, $N=1$, in the interior of the impedance tube. The translucent resin allows to see the array of HRs. Picture shows the tube open, but it was closed for the experiments. (b) Absorption of the system measured experimentally (crosses), calculated by the full modal expansion (thick continuous gray), effective parameters (dashed red), transfer matrix method (continuous blue) and finite element method (circles). (c) Representation of the reflection coefficient in the complex frequency plane for the optimized sample. Each line shows the trajectory of its zero by changing a geometry parameter. (d) Absorption peak as a function of the angle of incidence calculated by the effective parameters (dashed red), transfer matrix method (continuous blue). The inset in (d) shows the absorption coefficient in diffuse field as a function of frequency.}
	\label{fig:N1}
\end{figure*}

The previous sample provides perfect absorption for a thickness of $L=3a=\lambda/34.5$. Using an optimization method (sequential quadratic programming (SQP) method\cite{powell1978}) the geometry of the system can be tuned in order to minimize the thickness of the material, providing structures with perfect absorption and deep sub-wavelength dimensions. The TMM was employed in the optimization to consider the discreteness effects on the reflection coefficient. The resulting structure from the optimization procedure is shown in Figure.~\ref{fig:N1}~(a): a sample with a single layer of resonators, $N=1$ with $h=2.63$ mm, $d=14.9$ cm, $a= L = d/13 =1.1 $ cm, $w_n=2.25$ mm, $w_c=4.98$ mm, $l_n=2.31$ cm, $l_c=12.33$ cm. The width of the impedance tube used for measurements, $d$, allows to fit $13$ resonators in the transversal dimension as shown Fig.~\ref{fig:N1}~(a). The sample was built using stereolithography techniques using a photosensitive epoxy polymer (Accura 60{\textsuperscript{\textregistered}}, 3D Systems Corporation, Rock Hill, SC 29730, USA), where the acoustic properties of the solid phase are $\rho_0=1210$ kg/m$^3$, $c_0=[1570, 1690]$ m/s. The structure presents a peak of perfect absorption at $f=338.5$ Hz (different than that of the HR, $f_\mathrm{HR}=370$ Hz) with a thickness $L=\lambda/88$.

Figure~\ref{fig:N1}~(b) shows the absorption coefficient at normal incidence calculated with the different semi-analytical methods, predicted numerically by FEM and measured experimentally. At $f=338.5$ Hz, perfect absorption can be observed. The maximum absorption measured experimentally was $\alpha=0.97$, as shown in the inset of Fig.~\ref{fig:N1}~(b). This small discrepancy between the measurements and the models can be caused by experimental reasons including the non perfect fitting of the slit on the impedance tube and the excitation of plate modes of the solid medium that composes the metamaterial. 

On the other hand, Fig.~\ref{fig:N1}~(c) shows the corresponding reflection coefficient in the complex frequency plane calculated with MEM. The color map corresponds to the case in which the critical coupling condition is fulfilled, i.e., the zero of the reflection coefficient is exactly located on the real frequency axis. As long as the intrinsic losses depend on the geometry of the resonators and the thickness of the slits, we also represent in Fig.~\ref{fig:N1}~(c) the trajectory of this zero as the geometry of the system is modified. The crossing of the trajectories with the real frequency axis implies that perfect absorption can be achieved with this system at this particular frequency. It can be seen that the trajectories linked to the resonators geometry, $w_n$, $w_c$, $l_n$ have a strong effect in the real part of the complex frequency of the zero, as they modify the HRs resonant frequency. In the case of $l_n$, due to the geometric constraint $d\ge h + l_n+l_c$, increasing the length of the neck implies also the reduction of the cavity and the trajectory of the zero is twisted. The trajectory of the slit thickness, $h$, shows that the intrinsic losses are excessively increased for very narrow slits and critical coupling condition cannot be fulfilled. For very wide slits the geometrical constraints also imply the reduction of the size of the resonators and therefore the resonant frequency is increased. Finally, the trajectory linked to the lattice size, $a$, shows how the depth of the slit, $L=Na$, is mainly linked to the intrinsic losses of the system: the peak absorption frequency is almost independent of $a$, it mostly depends on the resonator resonant frequency. Moreover, as the slow-sound conditions are caused by the local resonance of the HRs, the periodicity of the array of HRs is not a necessary condition for these perfect absorbing panels. However, considering periodicity allows us to design and tune the system using the present analytical methods.

Finally, Fig.~\ref{fig:N1}~(d) shows the absorption of the metamaterial panel as a function of the angle of incidence. It can be observed that almost perfect absorption is obtained for a broad range of angles, being $\alpha>0.90$ for incident waves with $\theta<60^\circ$. The inset of Fig.~\ref{fig:N1}~(d) shows the absorption in diffuse field\cite{cox2009} calculated as $\alpha_{\mathrm{diff}}=2\int_{0}^{\pi/2}\alpha(\theta)\cos(\theta)\sin(\theta)\mathrm{d}\theta$, where at the working frequency it reaches a value of $\alpha_{\mathrm{di f f}}=0.93$, showing the omnidirectional behavior of the absorption in this sub-wavelength structure.

Realistic panels for sound perfect absorption with sub-wavelength sizes are designed in this work with simple structures made of bricks with Helmholtz resonators. Perfect absorption of sound is achieved at 338.5 Hz with a panel thickness of $L=\lambda/88 = 1.1$ cm and without added porous material. It is worth noting here that the total panel size in the vertical dimension is also sub-wavelength $d=\lambda/6.5=14.5$ cm. The sub-wavelength feature of the presented structure provides perfect absorption for a wide range of incident angles. This omnidirectional sound absorber can be employed in practical applications where the omnidirectional feature is mandatory. In addition, several theoretical approaches have been presented and validated experimentally, where their limits of validity are discussed. In order to provide accurate models, we have presented the design with standard Helmholtz resonators. However, the thickness of the structure can be even reduced by engineering the geometry using coiled-up channels or embedding the neck into the cavity of the HRs. These promising results open the possibilities to study different configurations based on these metamaterials and to extend the results to broadband and omnidirectional perfect absorption with deep sub-wavelength structures. 

\begin{acknowledgments}
	This work has been funded by the Metaudible project ANR-13-BS09-0003, co-funded by ANR and FRAE.
\end{acknowledgments}


%

\bigskip
\appendix
\section{Supplementary material to: \\
	Ultra-thin metamaterial for perfect and omnidirectional sound absorption}

The supplementary material for the paper entitled ``Ultra-thin metamaterial for perfect and omnidirectional sound absorption'' is presented. Firstly, the visco-thermal losses models are given. Secondly, details on the full modal expansion are presented. Finally, the modeling the Helmholtz resonators including the proper end corrections is presented. 

\subsection{Visco-thermal losses model}

\begin{figure}[b]
	\centering
	\includegraphics[width=8cm]{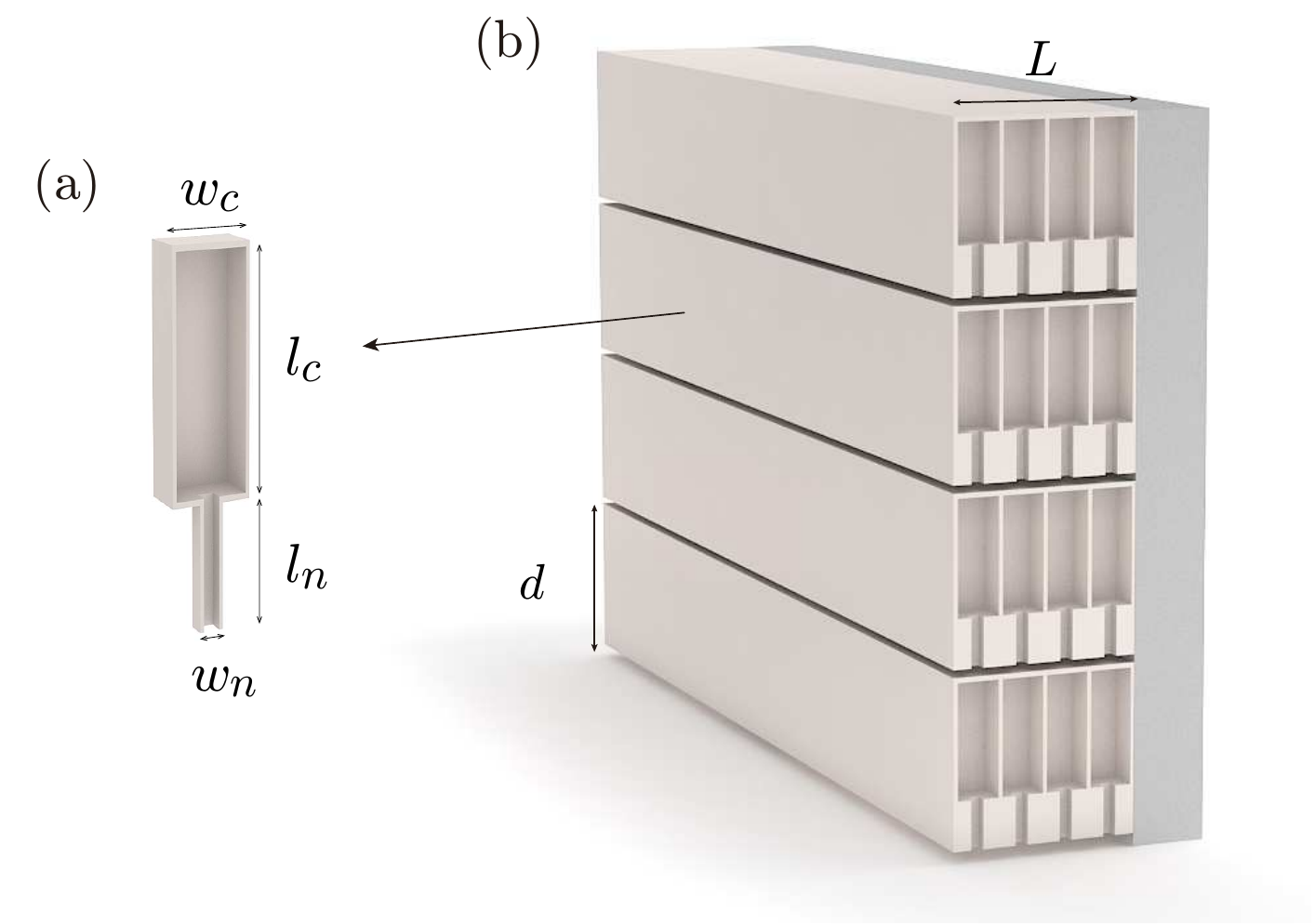}
	\caption{(a) Square cross-section Helmholtz resonator (HRs). (b) Conceptual view of the metamaterial panel placed on a rigid wall with $N=4$ layers of HRs.}
	\label{fig:FigSupp1}
\end{figure}

The system consist of a thin panel perforated with a periodic arrangement of closed slits, of thickness $h$, along the $x_1$ direction with periodicity $d$, as shown in Fig.~\ref{fig:FigSupp1}. The upper wall of the slit is loaded by $N$ identical Helmholtz resonators (HRs) in a square array of side $a$. We use HRs with square cross-section, characterized by a neck and cavity of with $w_n$, and $w_c$ and length $l_n$ and $l_c$ respectively. The visco-thermal losses in the system are considered both in the resonators and in the slit by using its effective complex and frequency dependent parameters \cite{stinson1991}.

{\bfseries{Slits}}: The effective parameters in the slit, considering only plane waves propagate inside, are expressed as:
\begin{equation}\label{eq:rhos}
\rho_s={\rho _0}\left[1-\frac{\tanh \left(\frac{h}{2}{G_\rho}\right)}{\frac{h}{2}{G_\rho}}\right]^{-1} \,,
\end{equation}
\begin{equation}\label{eq:Ks}
\kappa_s=\kappa_0\left[1+(\gamma-1)\frac{\tanh \left(\frac{h}{2}{G_\kappa}\right) }{\frac{h}{2}{G_\kappa}}\right]^{-1} \,,
\end{equation}
\noindent with $G_\rho=\sqrt{{\i\omega\rho_0}/{\eta}}$ and $G_\kappa=\sqrt{\i\omega\mathrm{Pr}\rho_0/{\eta}}$, and where $\gamma$ is the specific heat ratio of air, $P_0$ is the atmospheric pressure, $\Pr$ is the Prandtl number, $\eta$ the dynamic viscosity, $\rho_0$ the air density and $\kappa_0={\gamma P_0}$ the air bulk modulus.

{\bfseries{Ducts}}: The propagation in a rectangular cross-section tube can be described by its complex and frequency dependent density and bulk modulus, and considering that plane waves propagate inside, can be expressed as \cite{stinson1991}:
\begin{equation}\label{eq:rhoc}
\rho_t = -\frac{\rho_0 a^2 b^2}{4 G_\rho^2 \sum\limits_{k\in\mathbb{N}}\sum\limits_{m\in\mathbb{N}} \left[\alpha_k^2 \beta_m^2 \(\alpha_k^2 + \beta_m^2 - G_\rho^2\) \right]^{-1}} \,,
\end{equation}
\begin{equation}\label{eq:Kc}
\kappa_t=\frac{\kappa_0}{\gamma + \frac{4 (\gamma -1) G_\kappa^2 }{{a^2}{b^2}}\sum\limits_{k\in\mathbb{N}}\sum\limits_{m\in\mathbb{N}}{\left[\alpha _k^2\beta _m^2\(\alpha _k^2+\beta _m^2-G_\kappa^2\)\right]^{-1}}} \,,
\end{equation}
\noindent with the constants $\alpha _k=2(k+1/2)\pi/a$ and $\beta_m=2(m+1/2)\pi/b$, and the dimensions of the duct $a$ and $b$ being either the neck, $a=b=w_n$, or the cavity, $a=b=w_c$ of the Helmholtz resonators.

\subsection{Full modal expansion}

\begin{figure}[t]
	\centering
	\includegraphics[width=8cm]{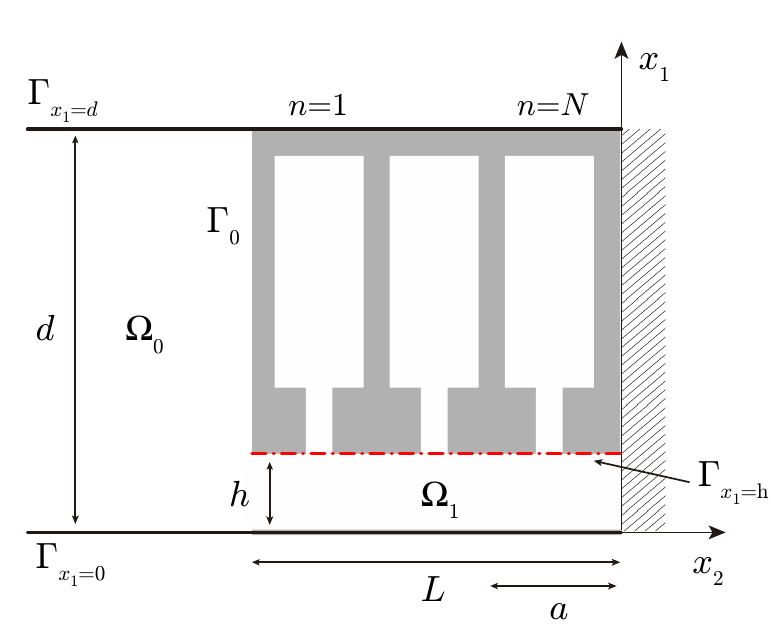}
	\caption{(a) Scheme of the unit cell of the panel composed of a set of $N$ Helmholtz resonators. Periodic boundary conditions are applied at boundaries $\Gamma_{x_1=d}$ and $\Gamma_{x_1=0}$, which reduces to symmetric (rigid) boundary conditions for normal incidence.}
	\label{fig:FigSupp2}
\end{figure}

As sketched in Fig.\ref{fig:FigSupp2}, the full space is divided into two sub-domains, the free air, $\Omega_0$, and the interior of the slit, $\Omega_1$. Periodic boundary conditions are assumed at the boundaries $\Gamma_{x_1=0}$ and $\Gamma_{x_1=d}$. It is worth noting here that for normal incidence periodic boundary condition reduce to symmetric (rigid) boundary conditions. At $\Gamma_{x_1=h}$ the effect of the resonators is included by a impedance condition given by $Z_\mathrm{wall} = Z_\mathrm{HR}/\phi$, with $Z_\mathrm{HR}$ the impedance of the HRs and $\phi=S_n^2/a^2$ surface porosity of the slit. 

First, the pressure field, $p^{(0)}$, in the domain $\mathrm{\Omega_0}$ is decomposed in $n$ modes as
\begin{equation}\label{eq:mp0}
p^{(0)}=\sum_{n=0}^{\infty} \cos \frac{n \pi}{d} x_1  \left\lbrace A_n^i \delta_n^0 e^{-ik_{2n}^{(0)} (x_2 - L)} + R_n e^{ik_{2n}^{(0)} (x_2 - L)} \right\rbrace ,
\end{equation}

\noindent where $A_n^i$ is the amplitude of the incident wave, $R_n$ the reflection coefficient of $n$ mode, $\delta_n^0$ the Kronecker delta,  
\begin{equation}
k_{2n}^{(0)} = \sqrt{k_0^2 - \left(\frac{n \pi }{d}\right)^2} \quad,\quad k_{1n}^{(0)}=\frac{n \pi }{d}\,,
\end{equation}
\noindent the wavenumbers inside the domain $\mathrm{\Omega_0}$ and in the direction $x_2$, and $x_1$, respectively, and $k_0$ the wavenumber in the air. The orthogonality relation for these modes reads as
\begin{eqnarray}
\int_0^d \cos\left(\frac{n \pi}{d} x_1\right) \cos \left(\frac{m \pi}{d} x_1\right) \: \mathrm{d} x_1 = \frac{\delta_{n}^{m} d}{\epsilon_m}, \label{project0}
\end{eqnarray}
\noindent where $\epsilon_m = 1$ if $m=0$ and $m = 2$ if $m\neq 0$. 

On the other hand, applying Neumann boundary conditions at $\Gamma_{x_1=0}$, $\Gamma_{x_1=d}$ and $\Gamma_{x_2=L}$, and the impedance condition at the wall $\Gamma_{x_1=h}$, the pressure $p^{(1)}$ in the domain $\mathrm{\Omega_1}$ can be decomposed as:
\begin{equation}\label{eq:mp1}
p^{(1)} = \sum_{m=0}^{\infty} A_m \cos\left(k_{2m}^{(1)} x_2\right) \cos \left(k_{1m}^{(1)} x_1\right) \,,
\end{equation}
where the wavenumber in the $x_2$ direction is
\begin{equation}\label{orth312}
k_{2m}^{(1)} = \sqrt{k_s^2 - \left(k_{1m}^{(1)}\right)^{2}},
\end{equation}

\noindent with $k_s=\omega (\rho_s/\kappa_s)^{1/2}$ the effective wavenumber in the slit. The wavenumber in the $x_1$ direction, $k^{(1)}_{1m}$, follows the dispersion relation \cite{redon2011}:
\begin{equation}\label{eqk1}
k_{1m}^{(1)} \tan\left (k_{1m}^{(1)} h\right) = \frac{- i \omega \rho_s}{Z_\mathrm{wall}}.
\end{equation}
This last equation is solved by the M\"{u}ller algorithm \cite{press2007} using the low-frequency initial value:
\begin{equation}
\hat{k}_{10}^{(1)} = \frac{1}{h} \sqrt{\frac{- \omega \rho_s h}{Z_\mathrm{wall}}}.
\end{equation}

The bi-orthogonality relation for the transversal modes $k_{1n}^{(1)}$ in the domain $\mathrm{\Omega_1}$ leads to:
\begin{eqnarray}\label{biorth1}
N_n=\frac{1}{h} \int_0^h \cos^2 k_{1n}^{(1)} x_1 \: \mathrm{d}x_1 =  \frac{1}{2}+ \frac{\sin 2 k_{1n}^{(1)} h}{4 k_{1n}^{(1)} h}  ,
\end{eqnarray}
if $\sin(2 k_{1n}^{(1)} h)+2 k_{1n}^{(1)} h \neq 0$.

Then, continuity boundary conditions at $x_2=L$ for pressure and the normal particle velocity are projected on the modes at the domains $\mathrm{\Omega_0}$ et $\mathrm{\Omega_1}$ on Eq.(\ref{eq:mp0}) and Eq.(\ref{eq:mp1}) respectively, leading finally to the linear system \cite{groby2015,groby2016}:
\begin{equation}\label{lineaire1}
\begin{split}
R_M - \dfrac{i \rho_0 \epsilon_M \phi_t}{\rho_s k_{2M}^{(0)}} \sum_{M'} R_{M'} \sum_n \dfrac{k_{2n}^{(0)} \tan(k_{2n}^{(1)}L) I_{nM} I_{nM'}}{N_n} =
& \\ A_M^i \delta_M^0 + \dfrac{i \rho_0 \epsilon_M \phi_t}{\rho_s k_{2M}^{(0)}} \sum_{M'} A_{M'}^i \delta_{M'}^0 \sum_n \dfrac{k_{2n}^{(0)} \tan(k_{2n}^{(1)}L) I_{nM} I_{nM'}}{N_n}.
\end{split}
\end{equation}
\noindent where $\phi_t=h/d$ is the total porosity and the integrals $I_{nm}^{(q)\pm}$:
\begin{eqnarray}
\begin{split}
I_{nm} =&\frac{1}{h} \int_0^h \cos\left(k_{1n}^{(1)} x_1\right) \cos\left(\frac{m \pi}{d} x_1\right) \: \mathrm{d}x_1  \\
=&  \dfrac{1}{2} \left\{ \text{sinc}\left [\left (k_{1n}^{(1)} + \frac{m \pi}{d}\right) h\right] + \text{sinc}\left[\left(k_{1n}^{(1)} - \frac{m \pi}{d}\right) h\right] \right \}.
\end{split}
\end{eqnarray}

Thus, the reflection coefficient for each mode, $R_M$, is calculated by this linear system of $M$ equations, Finally, the absorption of the metamaterial is calculated as 
\begin{equation}
\alpha= 1- \sum_n \frac{\mathrm{Re}\left(k_{2n}^{(0)}\right)\left|R_n\right|^2}{k_{20}^{(0)}}.
\end{equation}

On the other hand, using Eq.~(\ref{orth312}), the phase velocity in the slit can be calculated as:
\begin{eqnarray}
\hat{v}_{20}^{(1)}=c_s \left[1-\frac{\left(\hat{k}_{10}^{(1)}\right)^2}{k_s^2}\right]^{-1/2}\,,
\label{v20}
\end{eqnarray}
\noindent where $c_s=(\kappa_s/\rho_s)^{1/2}$.

\subsection{Transfer Matrix Method}
A discrete model is developed accounting for the finite number of resonators using the Transfer Matrix Method (TMM). Thus, for identical resonators, the transfer matrix is written as:
\begin{eqnarray}
\left( \begin{array}{c} 
{{P}_{{\rm i}}} \\ {{U}_{{\rm i}}} \end{array}\right)  
&=&\bf{T}
\left( \begin{array}{c} 
{{P}_{{\rm o}}} \\ 
{{U}_{{\rm o}}} 
\end{array} \right)\,,
\end{eqnarray}
\noindent where the transmission matrix $\bf{T}$ is written as
\begin{eqnarray}
{\bf{T}}=\left( \begin{array}{cc} 
{{T}_{11}} & {{T}_{12}} \\ 
{{T}_{21}} & {{T}_{22}}
\end{array} \right)
\nonumber 
=  {\bf M}_{\Delta l_{\mathrm{slit}}}({\bf M}_{s}{\bf M}_\mathrm{HR}{\bf M}_{s})^{N}\,.
\end{eqnarray}

\noindent Here, the transmission matrix for each lattice step in the slit, ${\bf M}_s$, is written as \renewcommand{\arraystretch}{2}
\begin{equation}
{{\bf M}_{s}}=\left( 
\begin{array}{cc} 
\cos\(k_s \dfrac{a}{2}\) & \i Z_s \sin \(k_s \dfrac{a}{2}\) \\ 
\dfrac{\i}{Z_s}\sin \(k_s \dfrac{a}{2}\) & \cos \(k_s \dfrac{a}{2}\) 
\end{array} \right).
\end{equation}
\noindent with $Z_s=({\kappa_s \rho_s})^{1/2}/S_s$ with $S_s=h\,a$. The resonators are introduced as a punctual scatters by a transmission matrix ${{\bf M}_{\mathrm{HR}}}$ as\renewcommand{\arraystretch}{1}
\begin{eqnarray}
{{\bf M}_{\mathrm{HR}}}=
\left( \begin{array}{cc} 
1 & 0 \\ 
1/{{Z}_{\mathrm{HR}}} & 1 
\end{array} \right),
\end{eqnarray}
\noindent and the radiation correction of the slit to the free space as 
\begin{eqnarray}
{\bf M}_{\Delta l_{\mathrm{slit}}}=
\left( \begin{array}{cc} 
1 & {Z}_{\Delta l_{\mathrm{slit}}} \\ 
0 & 1 
\end{array} \right),
\end{eqnarray}
\noindent with the characteristic radiation impedance ${Z}_{\Delta l_{\mathrm{slit}}}=-i\omega\Delta l_\mathrm{slit}\rho_0/\phi_t S_0$, where $S_0=d\,a$, $\rho_0$ the air density and $\Delta l_\mathrm{slit}$ the proper end correction that will be described later. 

Then, the reflection coefficient is calculated as
\begin{equation}
R=\frac{T_{11}-Z_0 T_{21}}{T_{11} + Z_0 T_{21}}.
\end{equation}
\noindent with $Z_0=\rho_0 c_0/S_0$, and finally the absorption as $\alpha=1-\left|R\right|^2$.

\subsection{Resonator impedance and end corrections}

Using the above effective parameters for the neck and cavity elements of a Helmholtz resonator, its impedance can be written as 
\begin{equation}
Z_\mathrm{HR}=i Z_n \frac{A-\tan k_n l_n \tan k_c l_c}{A \tan k_n l_n + \tan k_c l_c},
\end{equation}
\noindent with $A=Z_c /Z_n $, $l_n$ and $l_c$ are the neck and cavity lengths, $S_n=w_n^2$ and $S_c=w_c^2$ are the neck and cavity surfaces and $k_n$ and $k_c$, and $Z_n$ and $Z_c$ are the effective wavenumbers and effective characteristic impedance in the neck and cavity respectively.

It is worth noting here that this expression is not exact as long as correction due to the radiation should be included. The characteristic impedance accounting for the neck radiation can be expressed as:
\begin{widetext}
	\begin{equation}
	Z_\mathrm{HR} = -\i \frac{\cos(k_n l_n) \cos(k_c l_c) - Z_n k_n \Delta l \cos(k_n l_n) \sin(k_c l_c)/Z_c - Z_n \sin(k_n l_n)\sin(k_c l_c)/Z_c}{\sin(k_n l_n)\cos(k_c l_c)/Z_n - k_n \Delta l\sin(k_n l_n)\sin(k_c l_c)/Z_c + \cos(k_n l_n)\sin(k_c l_c)/Z_c}\,,
	\end{equation}
\end{widetext}
\noindent where the correction length is deduced from the addition of two correction lengths $\Delta l=\Delta l_1 + \Delta l_2$ as
\begin{eqnarray}
\Delta l_1 &=& 0.82 \left[1 - 1.35 \frac{r_n}{r_c} + 0.31 \(\frac{r_n}{r_c}\)^3\right] r_n \,,\\
\Delta l_2 &=& 0.82 \left[1 - 0.235 \frac{r_n}{r_s} - 1.32\(\frac{r_n}{r_t}\)^2 \right. \\
&&\left. + 1.54 \(\frac{r_n}{r_t}\)^3 - 0.86\(\frac{r_n}{r_t}\)^4 \right] r_n \,.
\end{eqnarray}

The first length correction, $\Delta l_1$, is due to pressure radiation at the discontinuity from the neck duct to the cavity of the Helmholtz resonator \cite{kergomard1987}, while the second $\Delta l_2$ comes from the radiation at the discontinuity from the neck to the principal waveguide \cite{dubos1999}. This correction only depends on the radius of the waveguides, so it becomes important when the duct length is comparable to the radius, i.e., for small neck lengths and for frequencies where $k r_n \ll 1$.

\begin{figure}[t]
	\centering
	\includegraphics[width=8.2cm]{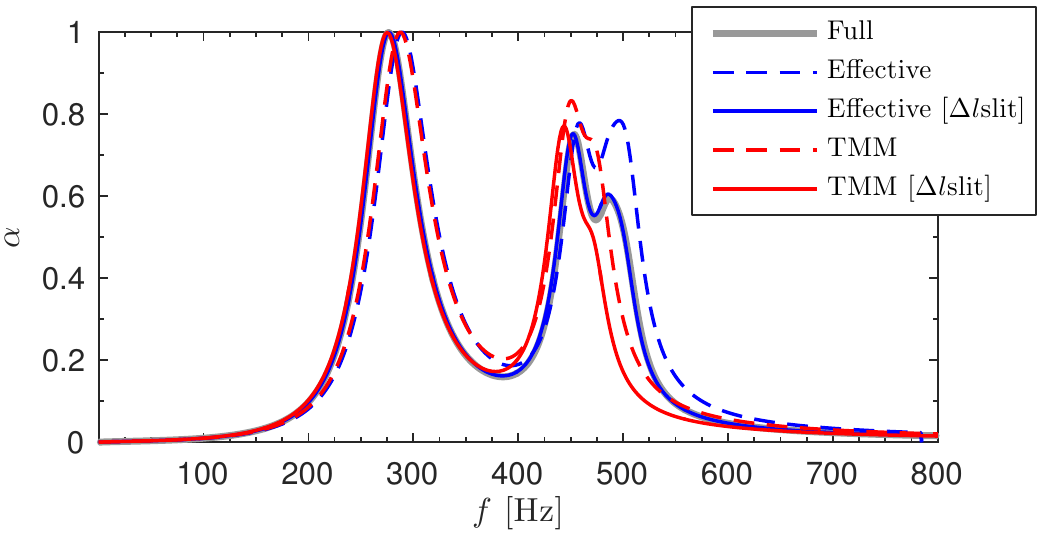}
	\caption{Absorption of a panel calculated with and without including the end correction of the slit for a panel of $N=3$ resonators with parameters $h=1.2$ mm, $a=1.2$ cm, $w_n=a/6$, $w_c=a/2$, $d=7$ cm, $l_n=d/3$, and $l_c=d-h-l_n$.}
	\label{fig:FigSupp3}
\end{figure}

Another important end correction comes from the radiation from the slits to the free air. The radiation correction for a periodic distribution of slits can be expressed as \cite{mechel2013}: 
\begin{equation}\label{eq:Dslit}
\Delta l_{\rm slit} = h \phi_t \sum_{n=1}^{\infty}\frac{\sin^2\(n\pi \phi_t\)}{(n\pi \phi_t)^3}.
\end{equation}

\noindent Note for $0.1\le \phi_t \le 0.7$ this expression reduces to $\Delta l_{\rm slit}\approx-{\sqrt{2}}\ln\left[ \sin\({\pi \phi_t}/{2}\)\right]/{\pi}$. Although Eq.~(\ref{eq:Dslit}) is appropriate for a periodic array of slits, it is not exact for slits loading HRs, therefore, we can evaluate a more realistic value for the end correction by reconstructing an equivalent impedance, $\tilde{Z}$, from the reflection coefficient of the zeroth order Bloch mode calculated with the full model and comparing it as \cite{groby2016}:
\begin{equation}\label{eq:Dslit2}
\tilde{Z} - \i Z_e \mathrm{cotan}(k_e L) = -i\omega\frac{\rho_0}{\phi_t} \Delta l_\mathrm{slit}
\end{equation}

The end correction using this last approach gives a value that depends on the geometry of the HRs and for the present examples is around 1.5 times the one using Eq.(\ref{eq:Dslit}). Figure~\ref{fig:FigSupp3} shows the absorption of the system using Eq.(\ref{eq:Dslit2}) and without any end correction. Perfect agreement between the full modal calculation and using the effective parameters can be obtained using the proper end correction. On the other hand, using the same length correction used in the TMM calculations perfect agreement is observed for the low frequency peaks of absorption curve.

\end{document}